\documentclass[prl,aps,twocolumn,reprint,floatfix,superscriptaddress,longbibliography,amssymb]{revtex4-1}
\usepackage[utf8]{inputenc} 
\usepackage{graphicx}
\usepackage{amsmath,amssymb}
\usepackage{accents}
\usepackage{float}
\newcommand{\be}{\begin{equation}}
\newcommand{\ee}{\end{equation}}
\newcommand{\bea}{\begin{eqnarray}}
\newcommand{\eea}{\end{eqnarray}}
\newcommand{\ben}{\begin{equation*}}
\newcommand{\een}{\end{equation*}}
\newcommand{\ba}{\begin{align}}
\newcommand{\ea}{\end{align}}

\newcommand{\mbf}{\mathbf}
\newcommand{\mrm}{\mathrm}
\newcommand{\su}{|\mathord{\uparrow}\rangle}
\newcommand{\sd}{|\mathord{\downarrow}\rangle}
\newcommand{\updownarrows}{\uparrow\mathrel{\mspace{-1mu}}\downarrow}
\newcommand{\downuparrows}{\downarrow\mathrel{\mspace{-1mu}}\uparrow}

\begin{document}

\title{Spinor self-ordering of a quantum gas in a cavity}

\author{Ronen M. Kroeze}
\altaffiliation[R.K.~and Y.G.~contributed equally to this work.]{}
\affiliation{Department of Physics, Stanford University, Stanford, CA 94305}
\affiliation{E.~L.~Ginzton Laboratory, Stanford University, Stanford, CA 94305}
\author{Yudan Guo}
\altaffiliation[R.K.~and Y.G.~contributed equally to this work.]{}
\affiliation{Department of Physics, Stanford University, Stanford, CA 94305}
\affiliation{E.~L.~Ginzton Laboratory, Stanford University, Stanford, CA 94305}
\author{Varun D. Vaidya}
\affiliation{Department of Physics, Stanford University, Stanford, CA 94305}
\affiliation{E.~L.~Ginzton Laboratory, Stanford University, Stanford, CA 94305}
\affiliation{Department of Applied Physics, Stanford University, Stanford, CA 94305}
\author{Jonathan Keeling}
\affiliation{SUPA, School of Physics and Astronomy, University of St Andrews, St Andrews KY16 9SS UK}
\author{Benjamin L. Lev}
\affiliation{Department of Physics, Stanford University, Stanford, CA 94305}
\affiliation{E.~L.~Ginzton Laboratory, Stanford University, Stanford, CA 94305}
\affiliation{Department of Applied Physics, Stanford University, Stanford, CA 94305}

\date{\today}

\begin{abstract}

We observe the joint spin-spatial (spinor) self-organization of a two-component BEC strongly coupled to an optical cavity. This unusual nonequilibrium Hepp-Lieb-Dicke phase transition is driven by an off-resonant two-photon Raman transition formed from  a classical pump field and the emergent quantum dynamical cavity field.  This mediates a spinor-spinor interaction that, above a critical strength, simultaneously organizes opposite spinor states of the BEC on opposite checkerboard configurations of an emergent 2D lattice. The resulting spinor density-wave polariton condensate is observed  by directly detecting the atomic spin and momentum state and by holographically reconstructing  the phase of the emitted cavity field. The latter provides a direct measure of the spin state, and a spin-spatial domain wall is observed. The photon-mediated spin interactions demonstrated here may be engineered to create dynamical gauge fields and quantum spin glasses.

\end{abstract}

\maketitle

The strong interaction between quantum matter and light provided by cavity quantum electrodynamics (QED) provides unique opportunities for exploring quantum many-body physics away from equilibrium~\cite{Ritsch2013,Sieberer:2016ej,Kirton:2018vv}. Discovering and classifying the properties of nonequilibrium quantum phase transitions is an active field~\cite{Diehl2010,Polkovnikov:2011iu,Sieberer2013}, with potential application to the engineering of quantum devices, such as those with superconducting correlations~\cite{Fausti:2011dy,Mitrano:2016fr}. One particularly rich setting in which to explore such physics is provided by systems realizing the driven-dissipative (Hepp-Lieb) Dicke model of two atomic states strongly coupled to an optical cavity field~\cite{Ritsch2013,Kirton:2018vv}. We present the observation of a nonequilibrium Dicke superradiant phase transition involving the spontaneous ordering of coupled atomic spin and spatial motion, as has been analyzed in Ref.~\cite{Mivehvar2017}.  The cavity photons mediate an effective position-dependent spin-spin interaction; the resulting transverse Ising model  that is realized opens future directions toward the study of artificial quantum spin glasses and neural networks in a driven-dissipative setting~\cite{Gopalakrishnan2011,Strack2011,Gopalakrishnan:2012cf,Buchhold:2013fr,McMahon:2016fy,Inagaki:2016eb,Torggler:2017hw,Rotondo18,Fiorelli:2018vf,Torggler:2018vu}. Moreover, with minor modification, this system could manifest dynamical gauge fields~\cite{Deng:2014gqa,Dong:2014cm,Padhi:2014go,Mivehvar:2014hr,Kollath:2016hs,Zheng:2016fa}, resulting in topological superfluids and exotic quantum Hall states.

As originally proposed~\cite{Dimer:2007da}, the nonequilibrium Dicke model describes an Ising ($Z_2$) symmetry-breaking transition of a spin-1/2 system coupled to a single cavity mode. The phase transition of the nonequilibrium Dicke model is closer to a classical than a quantum transition, though distinct from both~\cite{Nagy:2011fu,Torre:2013gc,Piazza:2013ip,Schutz:2015be,Kirton:2018vv}.  Experimentally, the nonequilibrium Dicke model could be realized by freezing the spins in a 2D lattice of period $\lambda/2$, where $\lambda$ is the wavelength of both the pump and cavity fields~\footnote{Note that the difference between the  cavity and pump frequencies is much smaller than their mean, and so we take $\lambda$ as the wavelength for both.}. The spins are disordered below the transition threshold and the cavity field is in a near-vacuum state. Above a pump threshold, the spins order in a $\lambda$-periodic checkerboard pattern (either up/down on the black/white sites or vice-versa) allowing the atoms to superradiantly scatter photons into the cavity mode. The emergent coherent field further orders the spins in a self-reinforcing manner.  Cavity dissipation stabilizes the driven, emergent spin order, and the phase of the cavity emission locks to either $0$ or $\pi$ relative to the pump phase depending on the symmetry-broken state.  Superradiant cavity emission of a spin-1 Dicke transition was observed with thermal atoms coupled to a cavity~\cite{Zhiqiang2017,Zhiqiang2018}.

Both pseudospin organization and superradiant emission have been observed in  an alternative form of the nonequilibrium Dicke transition~\cite{Nagy:2010dr,Baumann2010,Kessler2014}.  In this version, a Bose-Einstein condensate (BEC) matter wave is coupled to a cavity, where two different motional states play the role of up and down spin components. The atoms occupy either the black or white checkerboard sites (spaced $\lambda$-apart) of the emergent 2D lattice.  The pseudospin organization was detected by observing Bragg peaks at a momentum consistent with a checkerboard lattice together with detection of the relative phase locking of the pump and superradiant cavity emission~\cite{Baumann2010}.  The organized state may be called a `density-wave polariton condensate' in recognition of the joint light--matter-wave nature of the quasiparticles in the macroscopically occupied and coherent density-photon mode~\cite{Kollar2017}.  Roton instabilities and the extended Bose-Hubbard model have been realized~\cite{Mottl2012,Klinder2015,Landig2016}, and similar systems employing a few degenerate cavity modes have created a supersolid~\cite{Leonard2017}, an intertwined spatial order~\cite{Morales:2018gz}, and supermode-density-wave polariton condensates~\cite{Kollar2017}.  Highly degenerate cavities have been used to engineer tunable-range photon mediated atom-atom interactions~\cite{Vaidya:2018fp} that may lead to liquid crystalline states~\cite{Gopalakrishnan2009}.  A superradiant motional transition also occurs in cavities with spinless thermal atoms~\cite{Black2003,Domokos2002,Arnold2012}.  Self-organization of cold thermal gases and laser arrays due to optical feedback from a single mirror have also been observed~\cite{Nixon:2013fd,Pal:2017cd,Labeyrie2014,Robb2015,Labeyrie:2018uq}. 

\begin{figure}[t!]
\includegraphics[width = 0.49\textwidth]{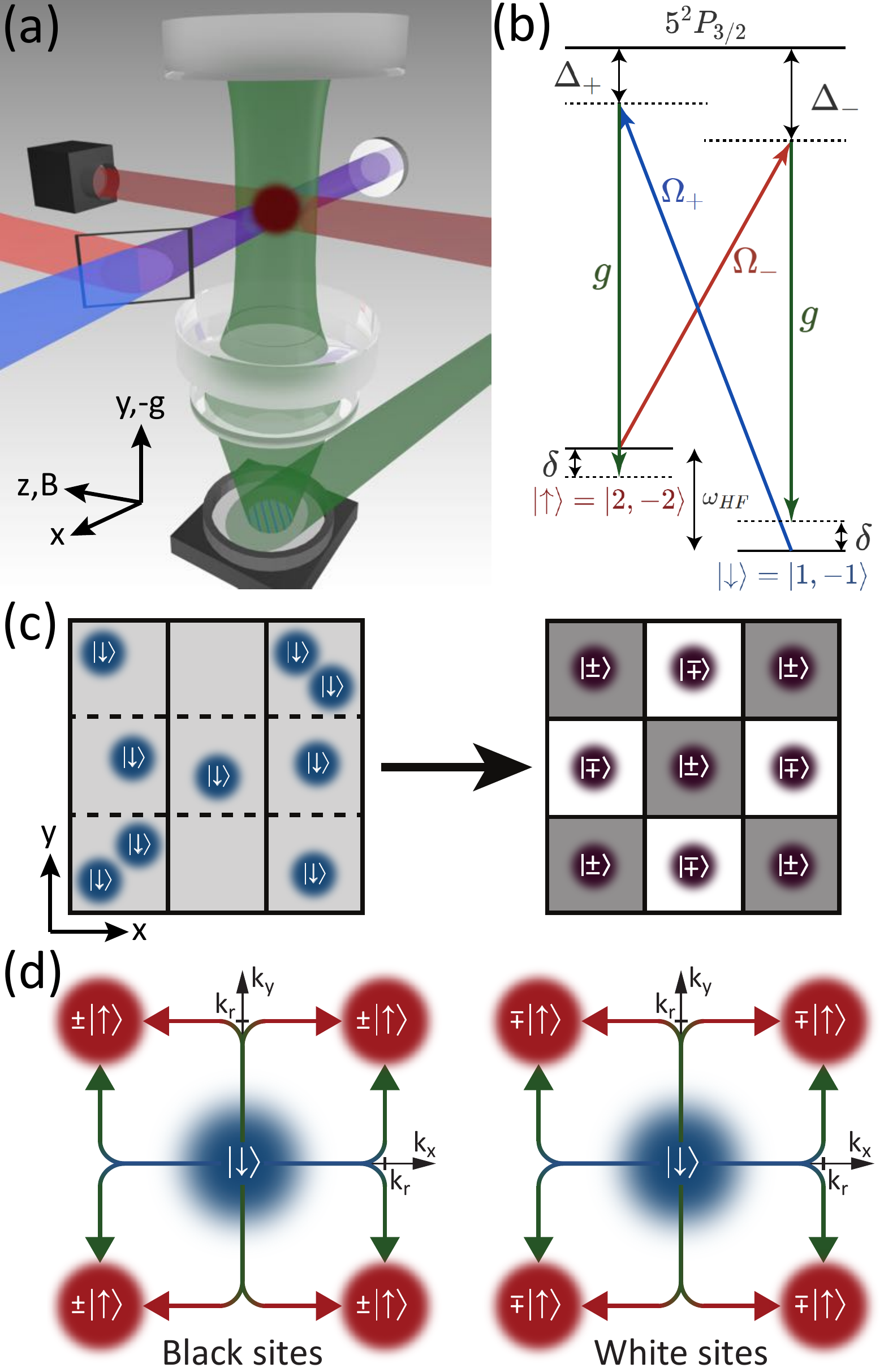}
\caption{(a) Experimental setup and detection techniques. The two Raman pump beams (red and blue), polarized along the cavity axis, are combined and retroreflected off the same mirror to create a phase-stable lattice (purple). The cavity mode (green), imaged onto a EMCCD camera, interferes with a local oscillator at an angle (also green). This provides the spatial heterodyne signal (blue lines) for the holographic reconstruction of the cavity field amplitude and phase. Momentum of the BEC (scarlet) is absorption-imaged in time-of-flight (scarlet beam). (b) Double Raman scheme for coupling two $^{87}$Rb Zeeman states, see text. (c) Real-space cartoon of the transition from randomly positioned atoms below threshold (left) to a checkerboard spinor order in an emergent 2D optical lattice above threshold (right). Atoms are in a $\hat{z}$ ($\hat{x}$) spin-polarization state below (above) threshold. Dashed (solid) lines in left panel are the nodes of the emergent cavity (pump) field. Solid lines in the right panel are the nodes of the above-threshold 2D optical lattice. (d) Momentum-space cartoons of the transition for atoms on black (left) versus white (right) sites. The state $|\pm,b\rangle$ ($|\mp,w\rangle$) emerges in black (white) checkerboard sites after sequential photon recoils from the pump and cavity fields.  Arrow colors depict the optical transition pathway shown in panel (b).} \label{experiment}
\end{figure}

What type of nonequilibrium phase transition arises when the pump and cavity fields  couple atomic motion and spin?  Reference~\cite{Mivehvar2017} describes such a system as a nonequilibrium spin-spatial Dicke superradiant phase transition in which atomic spins can flip while scattering photons into the cavity, picking up recoil momentum in the process~\footnote{The order-by-disorder aspect of the transition proposed in Ref.~\cite{Mivehvar2017} is not directly relevant to our system due to the initial spin polarization.}. This creates a spin-decorated checkerboard lattice, whose state is a `spinor density-wave-polariton condensate.'  The spinor density wave is described by the superposition of spinor operators $\hat{\psi}_{\uparrow,\downarrow}(\mathbf{r})$ described below, and arises due to a spinor-spinor interaction proportional to $\hat{\psi}^\dagger_{\uparrow}(\mbf{r}')\hat{\psi}^\dagger_{\downarrow}(\mbf{r})\hat{\psi}_{\downarrow}(\mbf{r}')\hat{\psi}_{\uparrow}(\mbf{r})$.  We note that this scenario is distinct from an emergent texture of a two-component BEC recently observed in a  miscible--immiscible transition created by a  state-dependent optical lattice arising from a nonequilibrium Dicke transition~\cite{Landini:2018bb}. In this experiment, the cavity mediated a density-density interaction $\rho_{+1}(\mbf{r})\rho_{-1}(\mbf{r'})$ between two Zeeman states $m=\pm1$ of a BEC and the two-component texture emerged above a critical ratio of the relative scalar and vector polarizabilities of the light fields.

We now describe the experimental system before reporting our observations of the superradiant spinor phase transition.  Figure~\ref{experiment}(a) shows the experimental configuration; see  previous work  for technical details of the cavity and the intracavity BEC production apparatus~\cite{Kollar2015,Vaidya:2018fp}. We trap within the  cavity a BEC of $4.1(3) \times 10^5$ $^{87}$Rb atoms in the $|F,m_F\rangle=|1,-1\rangle$ state. The BEC is confined in a crossed optical dipole trap (ODT) formed by a pair of $1064$-nm laser beams propagating along $\hat{x}$ and $\hat{z}$, resp. Using ODT shaping techniques~\cite{Bell2009}, we create a trap with frequencies $(\omega_x,\omega_y,\omega_z) = 2 \pi \times [58(1),63(1),47(1)]$~Hz that contains a BEC with Thomas-Fermi radii $(R_x, R_y, R_z) = [10.3(1), 9.4(1), 12.8(2)]$~$\mu$m.  These are all smaller than the $w_0 = 35$~$\mu$m waist of the TEM$_{0,0}$ cavity mode~\cite{siegman1986lasers}.  

%See Refs.~\cite{siegman1986lasers,Kollar2017,Supp} for discussion of cavity spectra.

To engineer the spinor Dicke Hamiltonian, we couple two internal states of $^{87}$Rb, $|F,m_F\rangle=|1,-1\rangle\equiv\sd$ and $|F,m_F\rangle=|2,-2\rangle\equiv\su$, through two cavity-assisted two-photon Raman processes; see Fig.~\ref{experiment}(b). A bias magnetic field of $\sim$2.83~G is applied along $+\hat{z}$, the direction of the quantization axis, resulting in an energy difference $\omega_\text{HF} \approx 6.829$ GHz between $\su$ and $\sd$ due to hyperfine splitting and Zeeman shifts. The Raman processes are created by the cavity and transversely oriented pump fields. The cavity field is that of the TEM$_{0,0}$ mode at frequency $\omega_c$ with coupling strength $g=g_0\Xi(x,z)$, where $g_0$ is the maximum single-atom coupling rate and $\Xi(x,z)$ is the transverse mode profile.  The pump beams have frequency $\omega_\pm$ such that $\omega_+ = \omega_- + 2(\omega_\text{HF}+ \delta)$, where $\delta$ is the two-photon Raman detuning.  Each pump field is far detuned from the atomic excited state by $\Delta_\pm$ with coupling strengths $\Omega_\pm$. Their mean frequency $\bar{\omega} = (\omega_++\omega_-)/2$ is detuned by $\Delta_c=\bar{\omega}-\omega_c$ from the cavity. The pump beams are retroreflected off the same mirror to create a phase-stable lattice. See Ref.~\cite{Supp} for a schematic of relative field frequencies, their generation and cavity spectra. 

\begin{figure}
\includegraphics[width = 0.45\textwidth]{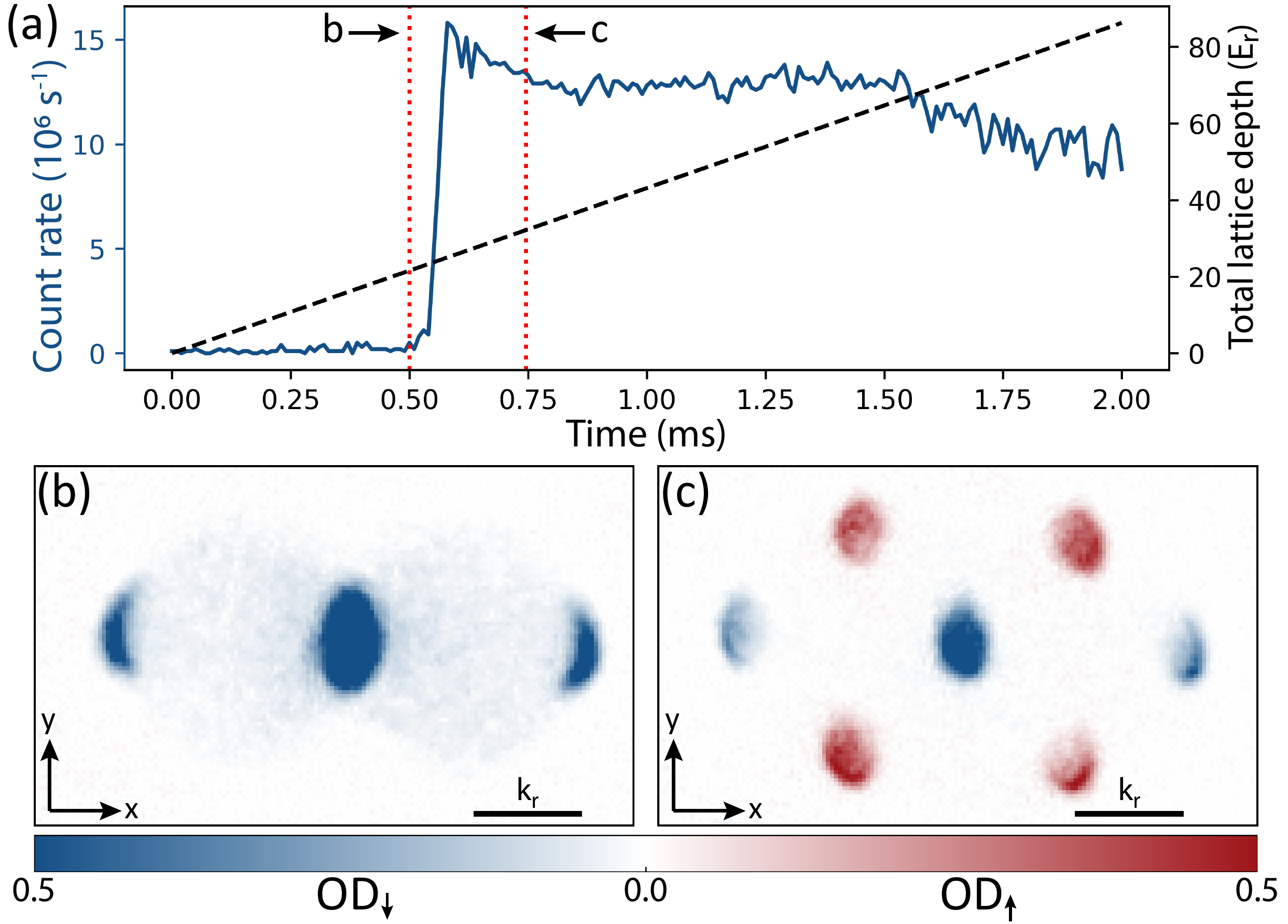}
\caption{(a) Cavity emission detected by single photon counters versus time plotted with the concomitant linear increase in lattice depth (proportional to pump intensity). The superradiant transition threshold is at $t\approx0.55$~ms. (b,c) Spin-sensitive absorption images of the atomic cloud in time-of-flight reveal the optical density (OD) of the momentum distribution of both spin states at the times indicated in panel (a). (b) All atoms are  in $\sd$ below threshold and either at zero momentum, or at $\mbf{k}=(\pm2k_r,0)$ due to  pump-lattice diffraction. (c) Above threshold, atoms have undergone a spin flip to $\su$ accompanied by a $\mbf{k}=\{(\pm k_r,\pm k_r);(\pm k_r,\mp k_r)\}$ momentum kick.  The resulting Bragg peaks are  spin-colored in the same pattern as in Fig.~\ref{experiment}(d).}\label{00superradiance}
\end{figure}

This coupling realizes the interaction Hamiltonian between the two components of the spinor state $\hat{\bm{\psi}}(\mbf{r}) = [\hat{\psi}_\uparrow(\mbf{r}), \hat{\psi}_\downarrow(\mbf{r})]^\intercal$ given by 
\be\label{int} 
H_\text{int}=\int d\mbf{r}\,2\eta\hat{\sigma}_x (\mbf{r}) (\hat{a}+\hat{a}^\dagger)\cos{k_rx}\cos{k_ry},
\ee
where the coupling strength $\eta$ is equal for both Raman transitions, $\hat{a}$ 
is the annihilation operator for the intracavity field, and  $\hat{\sigma}_x(\mbf{r}) = [\hat{\psi}^\dagger_\uparrow(\mbf{r})\hat{\psi}_\downarrow(\mbf{r}) + \hat{\psi}^\dagger_\downarrow(\mbf{r})\hat{\psi}_\uparrow(\mbf{r})]/2$. See Refs.~\cite{Supp,Mivehvar2017} for derivations and discussions of this model.
Given the initial state $\sd$, and  within the single recoil scattering limit~\footnote{This limit is valid under the condition of minimal depletion of the zero-momentum condensate; Fig.~\ref{00superradiance}(c) indeed shows that the $k=0$ peak is much more populous than the scattered momentum peaks.},  the spinor components take the form $\hat{\psi}_\downarrow(\mbf{r})=\hat{c}_\downarrow\psi_0(\mbf{r})$ and $\hat{\psi}_\uparrow(\mbf{r})=\hat{c}_\uparrow\psi_1(\mbf{r})$,  with the total atom number $N = \hat{c}_\uparrow^\dagger\hat{c}_\uparrow +\hat{c}_\downarrow^\dagger\hat{c}_\downarrow $.  The zero- and one-recoil  wavefunctions equal $\psi_0 =1$ and $\psi_1(\mbf{r}) = 2\cos{k_rx}\cos{k_ry}$, with the recoil momentum $\hbar k_r=2\pi\hbar/\lambda$. The form of $\psi_1(\mbf{r})$ is due to the 2D optical lattice emerging from the crossed pump and cavity standing-wave fields. 

Performing the spatial integral and defining pseudospin-1/2 operators as $\hat{J_z} = [\hat{c}_\uparrow^\dagger\hat{c}_\uparrow -\hat{c}_\downarrow^\dagger\hat{c}_\downarrow]/2$ and $\hat{J}_\pm=\hat{c}_{\updownarrows}^\dagger\hat{c}_{\downuparrows} $, we arrive at the spinor Dicke-model Hamiltonian~\cite{Supp}\footnote{We note that all parameters in this version of the Dicke model are tunable, providing access to a wealth of dynamical phenomena such as chaos and limit cycles~\cite{Emary:2003bd,Bhaseen:2012cq,Altland2012:qchaos,Altland2012:equil,Piazza2015,Kirton:2018vv,Zhiqiang2018}.}:
\be\label{dicke}
H_D=-\tilde{\Delta}_c\hat{a}^\dagger\hat{a} +(2\omega_r -\tilde{\delta})\hat{J}_z+\frac{\eta_D}{\sqrt{N}}(\hat{J}_++\hat{J}_-)(\hat{a}+\hat{a}^\dagger).
\ee
The $\hat{\mbf{J}}$ operate on the coupled pseudospin-1/2 spin-spatial degree of freedom.  The recoil frequency is $\omega_r = \hbar k_r^2/2m$, $\tilde{\Delta}_c$ is $\Delta_c$ minus the dispersive light shift, $\tilde{\delta} = \delta - \omega_s$, where  $\omega_s$ is the  ac Stark shift, and $\eta_D = \sqrt{N}\eta/2$. The first two terms account for the bare cavity energy and the energy shift between the spinor pseudospin states, resp.
  
The organized system exhibits a nonzero order parameter $\Theta \equiv \int d\mbf{r}\cos{k_rx}\cos{k_ry}\hat{\sigma}_x(\mbf{r})/N$ above a critical coupling strength $\eta_D>\eta_\text{th}$, where $\eta_\text{th} = [\tilde{\Delta}_c(2\omega_r-\tilde{\delta})]^{1/2}/2$ and $\Theta = \pm1$ in the $Z_2$-symmetry-broken state.   As shown in Fig.~\ref{experiment}(c), the organized state is one of the $|\pm,b\rangle + |\mp,w\rangle$ states of a spin-decorated $\lambda$-periodic checkerboard, where  $|\pm\rangle = \sd \pm \su$ are the $\hat{\sigma}_x$ eigenstates and $|b/w\rangle$ are the black/white checkerboard sites. The $Z_2$ broken-symmetry is reflected in the choice between $|+\rangle$ or $|-\rangle$ residing on black sites.  

Though staggered, the spinor pseudospin state is ferromagnetic.  This can be seen by integrating out the cavity field and rewriting Eq.~\ref{int} as an Ising Hamiltonian~\footnote{Writing Eq.~\ref{int} as an Ising Hamiltonian is possible because, unlike in Ref.~\cite{Zhiqiang2017}, we operate in the large-$\Delta_c$ dispersive regime.}:
\be H_\text{Ising}\propto \sum J_{ij}\cos{k_rx_i}\cos{k_rx_j}\cos{k_ry_i}\cos{k_ry_j}\hat{\sigma}^i_x\hat{\sigma}^j_x.
\ee  
The cosine terms can be incorporated into the $\hat{\sigma}_x$ through a local gauge rotation.  This results in a ferromagnetic, infinite-range ${J}_{ij}$ coupling of the locally rotated spin operators $\hat{\bar{\sigma}}^i_x$; see Ref.~\cite{Supp}. 

Figure~\ref{experiment}(d) presents the momentum-space cartoon of the transition. Above threshold, coherent two-photon Raman scattering creates a superposition of the atoms' initial zero-momentum-$\sd$ state and the $\pm\su$ state coupled to a momentum-recoil state comprised of the four superimposed $\mbf{k}=\{(\pm k_r,\pm k_r);(\pm k_r,\mp k_r)\}$ states~\footnote{Note, unlike in Refs.~\cite{Deng:2014gqa,Dong:2014cm,Padhi:2014go,Mivehvar:2014hr}, our Raman scheme does not produce (dynamical) spin-orbit-coupling  because this excited motional state has zero net momentum.}.  

We now present the observation of this organized spinor state in momentum space.
As in previous work~\cite{Baumann2010,Kollar2017,Zhiqiang2017}, superradiant cavity emission heralds the nonequilibrium Dicke phase transition; see Fig.~\ref{00superradiance}(a). We first demonstrate superradiance of the  model by linearly increasing the power in the Raman beams through the superradiant threshold with $\Delta_c = -4$~MHz and $\delta = -10$~kHz; see Ref.~\cite{Supp} for Raman-coupling-strength calibration~\footnote{This long-lived superradiance we observe is distinct from the single-beam spin-flip situation~\cite{Zhiqiang2018}.}. 

We then use spin-selective absorption imaging to detect the momentum distribution for each spin species independently during time-of-flight expansion of the gas. This method records the momentum of both spin components in a single realization of the experiment, allowing for observation of the spinor state associated with the spin-spatial self-ordering~\cite{Supp}. The spin dependent time-of-flight images are overlain in Figs.~\ref{00superradiance}(b) and (c)~\footnote{Spherical scattering halos are due to two-body contact interactions.}.   Below threshold, Fig.~\ref{00superradiance}(b) shows only $\sd$, zero-momentum atoms (and Bragg peaks from the pump lattice), while above threshold, Fig.~\ref{00superradiance}(c) shows that spin-decorated Bragg peaks appear in a fashion expected from Fig.~1(d). The absence of $\su$ atoms at $k=0$ and $\sd$ atoms at the 1st-order momentum peaks indicates that spinor order has emerged in the form of a $\lambda$-periodic checkerboard pattern in the $|\pm\rangle$ basis.

Above threshold, the frequency of the superradiant cavity emission should be locked  at $\bar{\omega}$~\cite{Dimer:2007da}.  Moreover, the phase of the emission should lock to either $0$ or $\pi$ (depending on the $Z_2$ broken-symmetry) with respect to a local oscillator (LO) field at $\omega_\text{LO}= \bar{\omega}+\delta_\text{LO}$. This  field is coherently generated from one of the  pump fields.  To establish that both effects occur, we  measure the  phase of the cavity field emission in a spatially resolved fashion using holographic reconstruction. Details of the  frequency stabilization and spatial heterodyne methods are in Ref.~\cite{Supp}. Briefly,  the LO is  shone at an angle onto the same EMCCD camera detecting the cavity emission, as depicted in Fig.~\ref{experiment}(a).  If the LO has the appropriate frequency (i.e., $\delta_\text{LO}=0$), the phase locking between the superradiant emission and the pump beam results in spatial interference fringes on the camera, realizing a spatial heterodyne measurement. Spatial Fourier demodulation analysis of the fringes reveals both the spatial dependence of the cavity field phase and amplitude~\cite{Supp}\footnote{See also Ref.~\cite{Schine:2018ui} for another recent use of this technique.}.

\begin{figure}
\includegraphics[width = 0.48\textwidth]{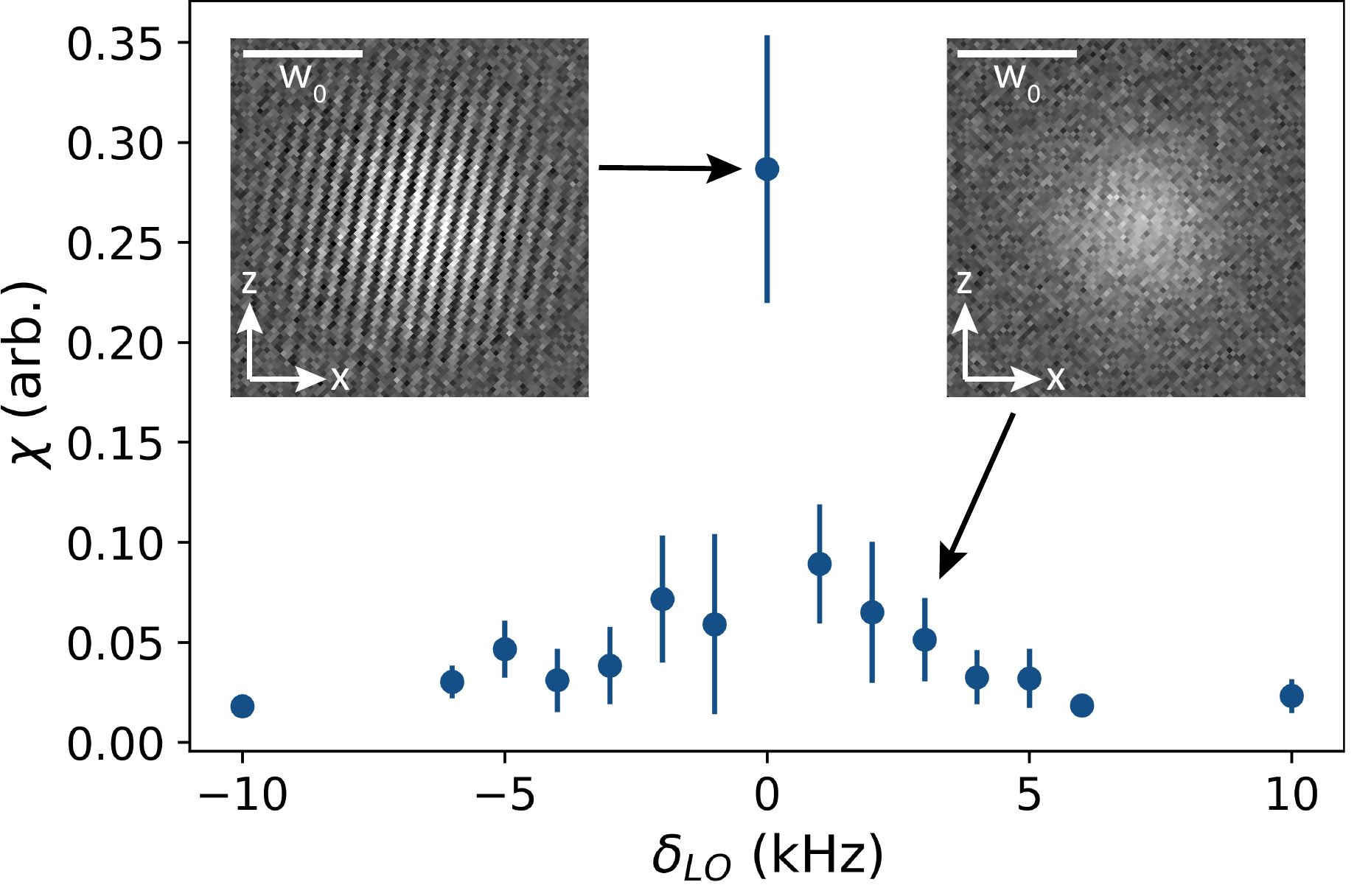}
\caption{Fringe amplitude factor $\chi$ as function of local oscillator frequency detuning $\delta_\text{LO}$.  The cavity is pumped above threshold  at a detuning  $\Delta_c = -4$~MHz from the TEM$_{0,0}$ cavity resonance.  The EMCCD camera integration time is 2~ms.   The insets show the spatial heterodyne signal---with local oscillator field subtracted for clarity---for both a maximal $\chi$ and at $\delta_\text{LO}= 3$~kHz where fringes average out. The error bars represent one standard deviation of the mean over five repetitions.}\label{00holography}
\end{figure}

The fringe amplitude factor $\chi$, defined in Ref.~\cite{Supp}, is plotted in Fig.~\ref{00holography}.  A distinct peak appears  at $\delta_\text{LO}=0$, as expected, while a significant averaging-out of  fringe contrast is manifest for detunings larger than $1/T$, where $T=2$~ms is the EMCCD integration time, due to a non-zero fringe phase velocity.  This demonstrates a unique feature of  the spinor Dicke model: cavity emission is detuned exactly halfway between the transverse pump beams, not at either or both of their frequencies. The high contrast fringes at $\delta_\text{LO}=0$ shows  that the phase is both stable and spatially constant over the superradiant emission pattern of the TEM$_{0,0}$ mode.

\begin{figure}
\includegraphics[width = 0.45\textwidth]{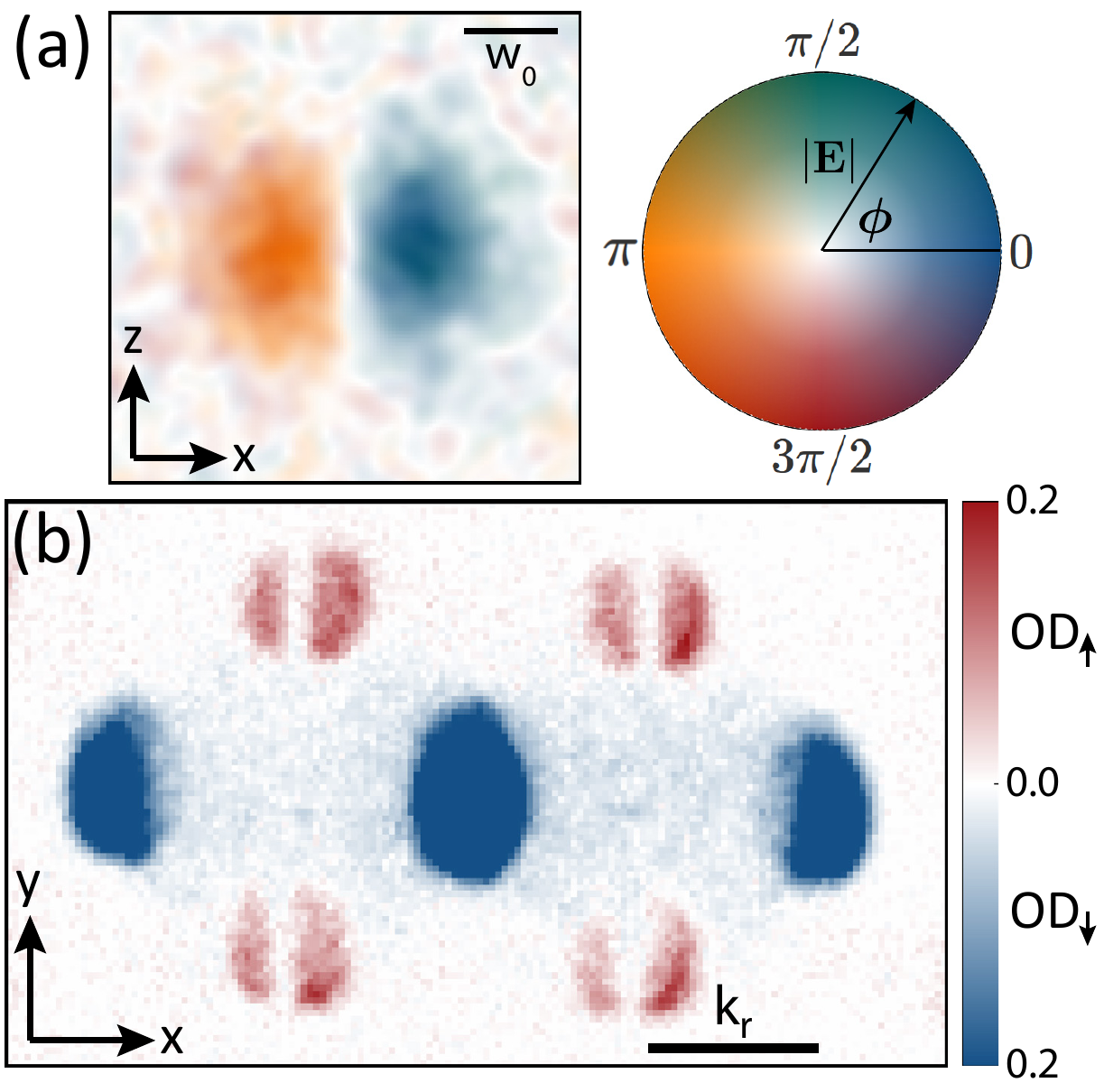}
\caption{(a) The cavity field amplitude and  phase, measured through holographic reconstruction, for a cavity locked near the  TEM$_{1,0}$  mode  whose  spatial profile $\Xi(x,z)$ exhibits a sign-flip at $x=0$. The phase of the right-hand lobe is defined as $0$ with respect to the local oscillator.  The phase shows a jump of exactly $\pi$ across the cavity center, demonstrating the fixed relative phase difference between the $\Theta=\pm1$ states with respect to the local oscillator phase.  (b) Observed spin-density structure factor. The small-$k$ transverse-mode-structure appears as a node in the 1st-order Bragg peaks. The combination of atomic and photonic observations indicates the existence of a domain wall in the spinor.} \label{01everything}
\end{figure}

We now present a measurement of the relative phase locking of the cavity and pump fields. This is determined both by observing a $\pi$ phase change of the superradiant emission across an induced spinor domain wall and by observing  a nodal structural factor in the 1st-order atomic Bragg peaks caused by this domain wall. To create adjacent spinor domains with opposite order parameter $\Theta$, the above experiment is repeated, but with the cavity frequency tuned near the 1st-order transverse mode TEM$_{1,0}$; $\bar{\omega}$ is set to $\Delta_c = -1$~MHz, see Ref.~\cite{Supp}. The field profile $\Xi(x,z)_{1,0}$ of this mode changes sign across the $x=0$ nodal line in the $x-z$ plane. The node appears in the superradiant cavity emission amplitude and phase are shown in Fig.~\ref{01everything}(a). The spinor order compensates for this sign change in the cavity field by flipping the  $Z_2$-symmetry-broken state from $\Theta = \pm1$ to $\mp1$ across the nodal line.  That is, the spin-spatial checkerboard pattern shifts by $\lambda/2$. The system does so to allow all the atoms to superradiantly emit into the cavity in phase, thereby minimizing the organization threshold.  This effect has been discussed for purely spatial organization~\cite{Kollar2017}. 

Holographic reconstruction of the emitted cavity field reveals the existence of this $\pi$ phase shift on either side of the nodal line; see Fig.~\ref{01everything}(a). The line defect also appears in the momentum distribution of the atoms shown in Fig.~\ref{01everything}(b).  A  node  in the 1st-order Bragg peaks appears due to the structure factor in the spinor organization~\cite{Kollar2017}. Together with the  phase flip of $\pi$,  the nodal
structure factor implies a spinor domain wall  along $(0,z)$.   In degenerate-mode cavities, such as the adjustable-length near-confocal cavity system of Refs.~\cite{Kollar2015,Kollar2017}, interference among modes could lead to topological spin-defect textures and local spin-spin interactions~\cite{Gopalakrishnan2011,Vaidya:2018fp}.

We have observed a spinor nonequilibrium Dicke superradiant phase transition among spinful atoms in a BEC coupled to a cavity. Moreover, the  intracavity photons mediate a spin-spin Ising interaction. This leads to a phase transition into a ferromagnetic state at a critical transverse field value associated with the two-photon pump intensity. By observing  both the photonic and atomic manifestations of the polaritonic system, we demonstrated joint spin-spatial self-organization. Using a higher-order transverse mode of the cavity and holographic reconstruction, we demonstrated the ability to create and image signatures of a domain wall. Strong Ising-type interactions, as realized here, may enable the study of quantum spin glass physics~\cite{Gopalakrishnan2011,Strack2011,Buchhold:2013fr}, which in turn may lead to quantum dissipative neuromorphic computing devices~\cite{Gopalakrishnan:2012cf,McMahon:2016fy,Inagaki:2016eb,Torggler:2017hw,Rotondo18,Fiorelli:2018vf,Torggler:2018vu}.  Lastly, a simple reconfiguration of the pump fields will enable the generation of dynamical spin-orbit coupling and gauge fields~\cite{Deng:2014gqa,Dong:2014cm,Padhi:2014go,Mivehvar:2014hr,Kollath:2016hs,Zheng:2016fa}.

We thank S.~Goldman and K.-Y.~Lin for experimental assistance and acknowledge funding support from the Army Research Office, the National Science Foundation under Grant No.~CCF-1640075, and by the Semiconductor Research Corporation under Grant No.~2016-EP-2693-C. J.~K. acknowledges support from SU2P.

%\bibliographystyle{apsrev4-1-prx}
%\bibliography{spinspatial_abbr}

%merlin.mbs apsrev4-1.bst 2010-07-25 4.21a (PWD, AO, DPC) hacked
%Control: key (0)
%Control: author (72) initials jnrlst
%Control: editor formatted (1) identically to author
%Control: production of article title (-1) disabled
%Control: page (0) single
%Control: year (1) truncated
%Control: production of eprint (0) enabled
%

\clearpage
\pagebreak

\section{Supplemental Material: \\ Spinor self-ordering of a quantum gas in a cavity}

\begin{figure}[b]
    \centering
    \includegraphics[width=0.48\textwidth]{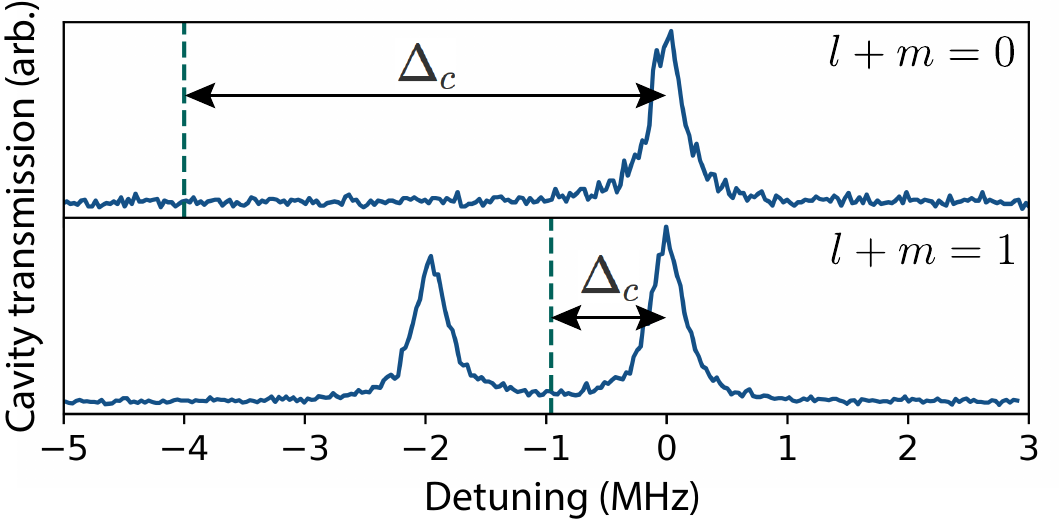}
    \caption{Transmission spectra of the cavity in use, at the $l+m=0$ family and the $l+m=1$ family. The lock points are indicated by dashed lines, giving $\Delta_c=-4$~MHz and $-0.96$~MHZ for the experiments involving TEM$_{0,0}$ and TEM$_{1,0}$ respectively.}
    \label{cavitytrans}
\end{figure}

\subsection{Cavity spectrum}
The length of our cavity can be adjusted in situ using a slip-stick piezo~\cite{Kollar2015}.  The length in this work is set such that the TEM$_{l,m}$ modes within $l+m=\text{const.}$ families~\cite{siegman1986lasers} are resolvable but far separated in frequency from other mode families.  Figure~\ref{cavitytrans} shows the cavity spectra for the two experiments discussed in this paper. For experiments using the TEM$_{0,0}$ mode, the cavity detuning is $\Delta_c=-4.00$~MHz, while $\tilde\Delta_c=-2.39$~MHz due to the dispersive shift (see the section on derivation of cavity-mediated spin-spin interaction below). Similarly, for the TEM$_{1,0}$ mode, $\Delta_c=-0.96$~MHz and $\tilde\Delta_c=-0.79$~MHz. That is, the detuning is blue of the TEM$_{0,1}$ mode, though red of the TEM$_{1,0}$ mode.  We observe dominant coupling to the TEM$_{1,0}$ mode and no instability from proximity to the blue of the TEM$_{1,0}$ mode. The splitting of approximately $50$~MHz between adjacent families of modes is at least an order of magnitude larger than these detunings.

\subsection{Frequency content}
\begin{figure}[b!]
    \centering
    \includegraphics[width=0.45\textwidth]{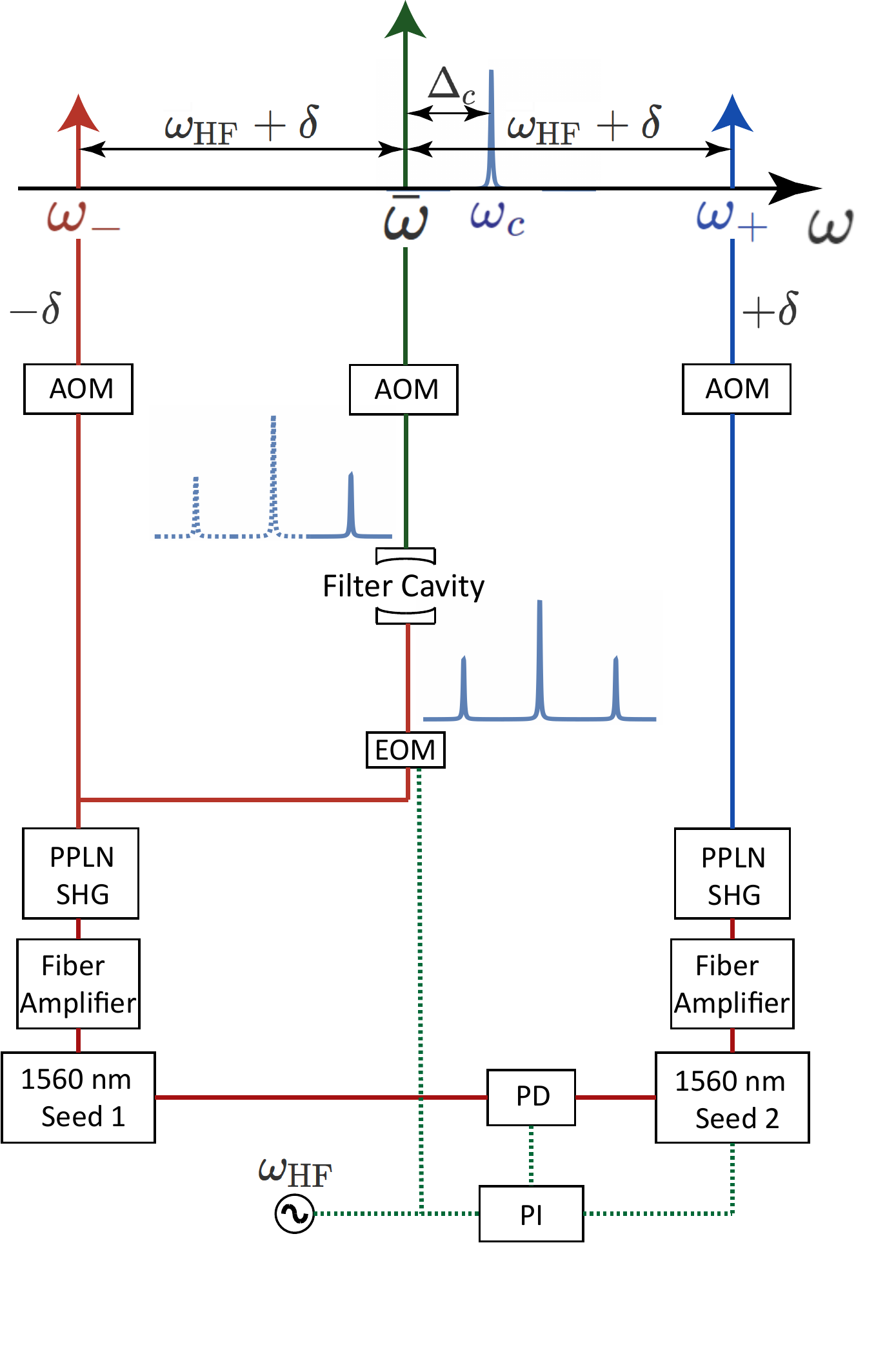}
    \caption{Schematic for the laser system used in this experiment. Green lines represent electrical signals. The two 780-nm Raman beams are derived using second harmonic generation (SHG) from two 1560-nm fiber lasers, whose relative frequency is stabilized at $\omega_\text{HF}$ with a beat-note lock referencing seed 2 to seed 1. After SHG, the frequencies of the two doubled light beams are separated by $2 \omega_\text{HF}$. AOMs placed in the path of the beams allow for additional frequency adjustments and intensity control. The science cavity is stabilized at $\omega_c$ using 1560-nm light from seed 1 through the Pound-Drever-Hall (PDH) technique. The same rf source used to lock the fiber lasers is used to drive an electro-optic modulator (EOM) for the purpose of generating the local oscillator beam at $\omega_\text{LO}$. The correct sideband is isolated by a filter cavity.}
    \label{frequency}
\end{figure}
The frequency content of the laser beams is schematically summarized in Fig.~\ref{frequency}. Both 780-nm Raman beams are derived from frequency-doubled 1560-nm light. The relative frequency between the two 1560-nm seed lasers are stabilized with respect to a stable frequency source calibrated via microwave spectroscopy to oscillate at the frequency difference $\omega_\text{HF}$ between $|1,-1\rangle$ and $|2,-2\rangle$.  $\omega_\text{HF}$ includes the Zeeman shift associated with the applied magnetic field. This frequency difference is controlled using a proportional-integral loop filter with feedback applied on seed 2. Additional 1560-nm light from seed 1 is used to stabilize the science cavity using the Pound-Drever-Hall (PDH) technique. The cavity resonance frequency $\omega_c$ is detuned from the 5$^{2}S_{1/2}|2,-2\rangle$ to 5$^{2}P_{3/2}$ transition by 154 GHz. The resulting Raman pump beams have atomic detunings of $160$ and $147$~GHz, resp. Using a fiber electro-optic modulator (EOM), sidebands at $\omega_\text{HF}$ are added onto the $\omega_-$ Raman beam in a separate path to derive the cavity probe (for use in taking the data in Fig.~\ref{cavitytrans}) and local oscillator beam. The drive signal to the EOM is split off from the same source that locks the seed lasers.  We further isolate the correct sideband from the EOM output using a filter cavity; the resulting beam is at the mean frequency $\bar\omega$ of the two Raman beams and phase stable with respect to the cavity. Additional acousto-optic modulators (AOMs) provide intensity stabilization and additional frequency-shifting capabilities to symmetrically adjust the Raman detuning $\delta$. All rf signals used in the experiment are stabilized with respect to the same 10-MHz Rb clock.

\subsection{Holographic reconstruction}
\begin{figure}[b!]
    \centering
    \includegraphics[width=0.47\textwidth]{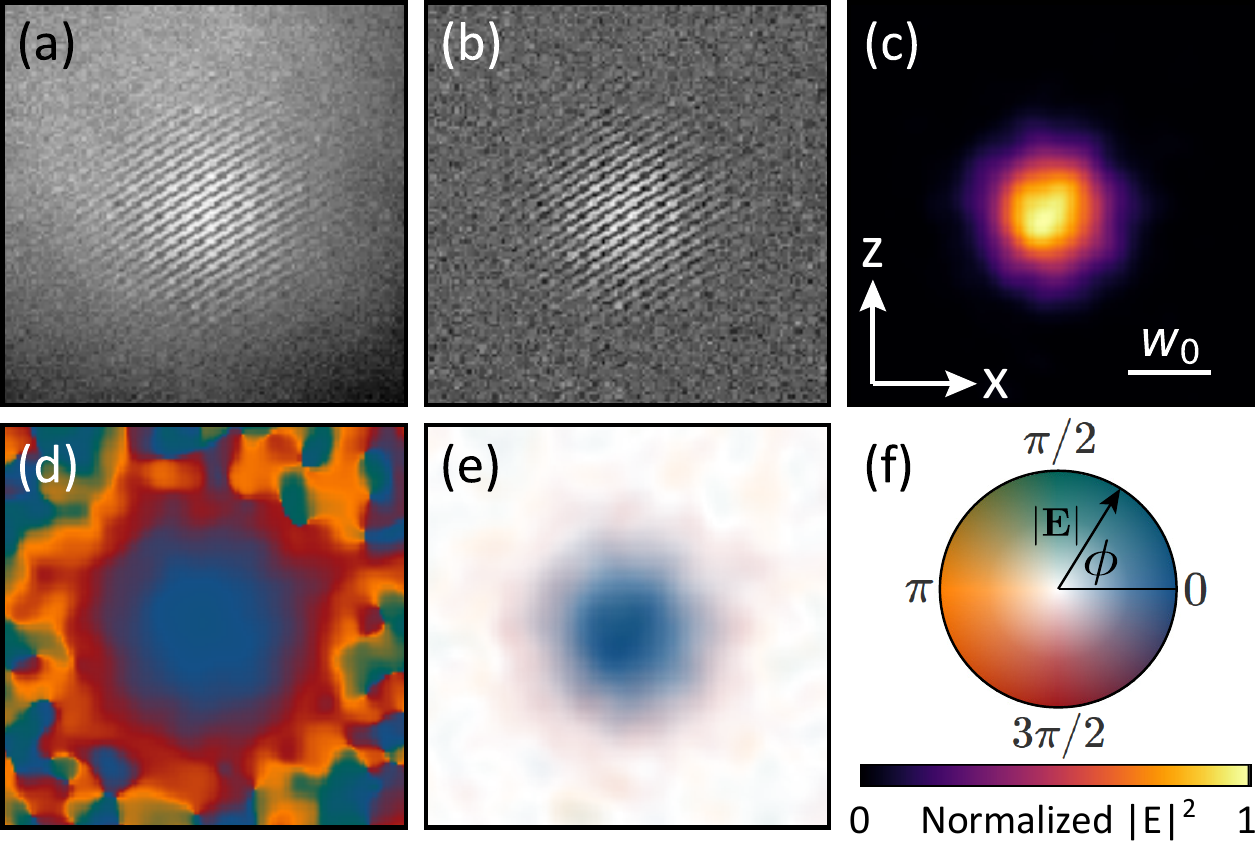}
    \caption{Holographic reconstruction of cavity fields. (a) Camera image $I_h(\mbf{r})$ generated by the interference between cavity emission from a TEM$_\text{0,0}$ mode and the local oscillator beam. (b) The corrected image $E_\text{corr}(\mbf{r})$ generated from (a) using the procedure described in Eq.~\ref{holo_correction}. (c) The intensity $|E_c(\mbf{r})|^2$ of the cavity emission extracted from  (b). (d) The phase of the cavity emission extracted from (b).  (e) A visualization of the complex electric field constructed using the amplitude from (c) and phase from (d). (f) Color legend for panels (c) and (e).  Color wheel is for panel (e) while color bar is for panel (c).}
    \label{holograpy}
\end{figure}

Above threshold, the superradiant cavity emission at frequency $\bar{\omega}$ observed on our EMCCD camera can have both amplitude and phase fluctuations in space. In the most general case, this field may be expressed as $E_c(\mbf{r}) = |E_c(\mbf{r})|e^{i \phi_c(\mbf{r})}$. The amplitude and phase of this field is measured using a holographic technique; see Ref.~\cite{Schine:2018ui} for another recent demonstration of this technique. A large local oscillator (LO) beam at frequency $\bar{\omega}+\delta_\text{LO}$ is incident on the EMCCD camera with a wavevector $\Delta \mbf{k}$ relative to the cavity emission. This LO beam is derived from the output of the filter cavity in Fig.~\ref{frequency} and the AOM provides a controllable frequency shift $\delta_\text{LO}$.  The interference between the cavity emission $E_c(\mbf{r})$ and the LO field $E_\text{LO}(\mbf{r})$ produces an image with an intensity $I_h(\mbf{r})$ on the EMCCD camera; see Fig.~\ref{holograpy}(a). This may be expressed as
\begin{widetext}
\begin{align}
I_{h}(\mbf{r}) = \lvert E_c(\mbf{r}) \rvert ^2 + \lvert E_\text{LO}(\mbf{r}) \rvert ^2 + 2\chi(\delta_\text{LO})|E_c(\mbf{r})E_\text{LO}(\mbf{r})| \cos \left( \Delta \mbf{k} \cdot \mbf{r} + \Delta\phi(\mathbf{r}) \right),
\label{hologram}
\end{align}
\end{widetext}
where $\Delta\phi(\mathbf{r})= \phi_c(\mathbf{r}) - \phi_\text{LO}(\mathbf{r})$ is the phase difference between the cavity and LO wavefronts. Both $|E_c(\mbf{r})|$ and $\phi_c(\mbf{r})$ are inferred from the amplitude and phase of the fringes produced by the oscillatory term of Eq.~\ref{hologram}. Reduction of fringe contrast is characterized by the factor $\chi(\delta_\text{LO})$. Several factors contribute to this reduction.  For example, mismatch in the spatial and polarization-mode overlap of the cavity and LO reduces contrast.  The contrast can also appear smaller due to a frequency difference between the LO and cavity emission: the fringe signals spatially average during the EMCCD camera's 2-ms integration time because the fringes have a non-zero phase velocity.  This spatial averaging effect allows us to determine the cavity emission via measuring fringe contrast versus $\delta_\text{LO}$, as shown in Fig.~\ref{00holography}. Noise in the relative frequency between the cavity emission and LO also leads to spatial averaging.  

In order to accurately extract $|E_c(\mbf{r})|$ and $\phi_c(\mbf{r})$, the image must first be corrected to account for intensity and phase variations of the LO beam. An independent measurement of the local oscillator intensity $I_\text{LO}(\mbf{r}) = \lvert E_\text{LO}(\mbf{r}) \rvert^2$ allows us to create a corrected field image $E_{\mrm{corr}}(\mbf{r})$ whose fringe amplitude solely depends on $|E_c(\mbf{r})|$:
\begin{align}
&E_{\text{corr}}(\mbf{r}) = \frac{I_{h}(\mbf{r}) - I_\text{LO}(\mbf{r})}{\sqrt{I_\text{LO}(\mbf{r})}}\nonumber \\
&= \frac{|E_c(\mbf{r})|^2}{|E_\text{LO}(\mbf{r})|} + 2\chi(\delta_\text{LO})|E_c(\mbf{r})|\cos \left( \Delta \mbf{k} \cdot \mbf{r} + \Delta\phi(\mathbf{r}) \right).
\label{holo_correction}
\end{align}
See Fig.~\ref{holograpy}(b) for plot of $E_{\text{corr}}(\mbf{r})$.
Assuming the cavity field varies slowly over the fringe wavelength $2\pi/|\Delta \mathbf{k}|$, we may extract $|E_c(\mbf{r})|$, shown in Fig.~\ref{holograpy}(c), and $\Delta \phi(\mbf{r})$, shown in Fig.~\ref{holograpy}(d), by demodulating $E_{\mrm{corr}}$ at the fringe wavevector $\Delta \mbf{k}$.

Finally, the phase of the cavity field may be extracted from $\Delta \phi(\mbf{r})$ by correcting for phase variations $\phi_\text{LO}(\mbf{r})$ of the local oscillator wavefront. The TEM$_\text{0,0}$ mode of the cavity is used to calibrate these variations since it has a uniform phase over its transverse profile. Measuring $\phi_\text{LO}(\mbf{r})$ in this manner allows us to calculate the phase of the cavity wavefront as $\phi_c = \Delta \phi + \phi_\text{LO}$ and consequently visualize the complex electric field of higher-order modes as shown in Fig.~\ref{01everything}(b).

\subsection{Spin-selective imaging}
At the end of the experimental sequence, atoms can be in either $\sd$, $\su$, or a superposition of the two. To selectively detect these states, we perform absorption imaging on the cycling transition between 5$^{2}S_{1/2}|2,-2\rangle$ and 5$^{2}P_{3/2}|3,-3\rangle$. Only the atoms in $\su$ are imaged due to the absence of repumping light. We verified that there is negligible depumping with circularly polarized light that drives purely $\sigma_-$ transitions. Following this initial imaging pulse, an intense pulse of light resonant with the transition is applied, which results in the expulsion of the $\su$ population from view. Next, the atoms in $\sd$ are transferred to 5$^{2}S_{1/2}|2,-2\rangle$ using microwave adiabatic rapid passage and imaged using the same cycling transition. These atoms are subsequently also removed from the field of view, after which `bright' and `dark' images are taken for completing the absorption imaging process. The extracted optical densities from the first and second imaging pulse are then  overlaid to produce spin-full absorption images such as those presented in Figs.~\ref{00superradiance}(b), (c) and~\ref{01everything}(b).

\subsection{Derivation of cavity mediated spin-spin interaction}\label{model}
The Hamiltonian of a single cavity mode $a$ with spatial profile $\Xi(\mbf{r})$ interacting with atoms can be written as in Ref.~\cite{Zhiqiang2018}:
\be\label{basicH}
H = \omega_c \hat{a}^{\dagger} \hat{a} + H_{\mathrm{atom}}+H_{\mathrm{trap}} + H_{\mathrm{kinetic}} + H_{\mrm{int}},
\ee
where $\omega_c$ is the optical frequency of the cavity mode, $H_{\mathrm{atom}}$ is the energy of the atomic internal states, and $H_{\mathrm{trap}}$ and $H_{\mathrm{kinetic}}$ capture the potential and kinetic energy of atoms in different internal states. $H_{\mrm{int}}$ describes the coupling introduced by the pump beams (with optical frequency $\omega_+$ and $\omega_-$) and cavity:
\begin{align}
H_{\mrm{int}} &= \int d^3 \mbf{r} \frac{1}{\sqrt{2}} \Big( \Omega_{+} (\mbf{r})e^{-i \omega_{+} t} + \Omega_{-}(\mbf{r}) e^{-i \omega_{-} t} \Big)  \nonumber \\
&\times \displaystyle\sum_{FF'} \Big( \hat{A}^{+1}_{FF'} (\mbf{r}) - \hat{A}^{-1}_{FF'} (\mbf{r}) \Big)  \nonumber \\ &+\int d^3 g_0 \Xi(\mbf{r}) \hat{a}  \displaystyle\sum_{FF'} \hat{A}^{0}_{FF'} (\mbf{r}) + \text{H.c.},
\end{align}
where
\be
\hat{A}^{(q)}_{FF'} (\mbf{r}) = \displaystyle\sum_{m} c(F,m \rightarrow F', m+q) \hat{\psi}^{\dagger}_{F',m+q} (\mbf{r}) \hat{\psi}^{~}_{F,m} (\mbf{r})
\ee
is the atomic raising operators connecting different hyperfine levels of the ground $\hat{\psi}^{~}_{F,m}$ and excited $\hat{\psi}_{F',m+q}$ states.  The   Clebsch-Gordon coefficients  $c(F,m \rightarrow F', m+q)$ are the relative strengths of the transitions. We apply a bias magnetic field along $\hat{z}$. Both pump beams are linearly polarized along the cavity axis. The additional factor of $1/\sqrt{2}$ for the Rabi frequency of the two pump beams $\Omega_{+}$ and $\Omega_-$ comes from the fact that the beams couple to both $\sigma_{+}$ and $\sigma_{-}$ transitions, though only one is close to resonant for each beam due to Zeeman shifts. 

The spatial profile of mode $\Xi$ results in a spatially dependent single-photon Rabi frequency $g_0 \Xi(\mbf{r})/\Xi_{0,0}(0)$, where $\Xi_{0,0}$ is the profile of a TEM$_{0,0}$ mode. Given the  large detunings of the pumps from the atomic excited states compared to the excited-state hyperfine splittings, all the excited states are assumed to be at the same energy $\omega_a$. In the ground states, the Zeeman shift pushes $|F=2,m_F=0\rangle$ out of resonance, so we only consider the spin components $|F,m_F\rangle=|1,-1\rangle\equiv\sd$ and $|F,m_F\rangle=|2,-2\rangle\equiv\su$ of the atom's hyperfine states as the coupled two-level system. All energy levels are defined with respect to the energy of $\sd$, and the bare energy splitting $\omega_\text{HF}$ (hyperfine splitting plus additional Zeeman shift) between $\su$ and $\sd$ is set by the bias magnetic field along $\hat{z}$ of $\sim 2.82~\mrm{G}$. We use microwave spectroscopy to calibrate the field and estimate a field fluctuation-induced frequency noise of 2.4~kHz on $\omega_\text{HF}$.

To obtain the effective Hamiltonian, we transform Eq.~\ref{basicH} into a rotating frame defined by the unitary transformation $\hat{U} = \mathrm{exp} (-i \hat{H}_{t} t)$, where
\be
\hat{H}_{t} = \frac{1}{2} (\omega_+ + \omega_-) \hat{a}^{\dagger} \hat{a} + \frac{1}{2} (\omega_+ - \omega_-) \int d^3 \mbf{r}\, \hat{\psi}^{\dagger}_{\uparrow}(\mbf{r}) \hat{\psi}_{\uparrow}(\mbf{r}).
\ee
Here, the coupled spin-spatial atomic states are represented by the spinor $\hat{\bm{\psi}}(\mbf{r}) = [\hat{\psi}_\uparrow(\mbf{r}), \hat{\psi}_\downarrow(\mbf{r})]^\intercal$.
Before writing the resulting Hamiltonian, we define the detunings $\Delta_+$ and $\Delta_-$ from the atomic excited state for each of the Raman transitions, the detuning $\Delta_c$ of the mean pump frequency from the cavity frequency $\omega_c$, and the two-photon detuning of the cavity-assisted Raman transition resonance $\delta$ as:
\begin{align}
\Delta_+ &= \omega_+ - \omega_a \nonumber \\
\Delta_- &= \omega_- + \omega_\text{HF} - \omega_a \nonumber \\
\bar{\omega} &= \frac{1}{2} (\omega_+ + \omega_-) \nonumber \\
\Delta_c &= \frac{1}{2} (\omega_+ + \omega_-) - \omega_c \nonumber \\
\delta &= \frac{1}{2} (\omega_+ - \omega_-) - \omega_\text{HF}.
\end{align}
We set $\delta \approx-10$ kHz, while the detuning for  other allowed Raman processes, e.g., the coupling between $\sd$ and $|F=2, m_F=0\rangle$, is on the order of a few MHz due to Zeeman splitting. After adiabatically eliminating the atomic excited states and ignoring the $s$-wave interaction and external harmonic trapping potential, the resulting Hamiltonian $H = H_{\uparrow} + H_{\downarrow} + H_{\mathrm{cavity}} + H_{\mathrm{Raman}}$ in the rotating frame is given by
\begin{widetext}
\begin{align}
H_{\uparrow} &= \int d^3 \mbf{r} \hat{\psi}^{\dagger}_{\uparrow}(\mbf{r}) \Big[ -\frac{\hat{\mbf{p}}^2}{2 m}  + \bigg( \frac{\Omega^2_{+}}{6(\Delta_+ + \omega_\text{HF})} +  \frac{\Omega^2_{-}}{6\Delta_-} \bigg) \cos^2(k_r x) + \frac{[g_0 \Xi(x,z)]^2}{\Delta_+} \cos^2(k_r y) \hat{a}^{\dagger} \hat{a} - \delta \Big] \hat{\psi}_{\uparrow} (\mbf{r}) \nonumber \\
H_{\downarrow} &= \int d^3 \mbf{r} \hat{\psi}^{\dagger}_{\downarrow}(\mbf{r}) \Big[ -\frac{\hat{\mbf{p}}^2}{2 m} + \bigg( \frac{\Omega^2_{+}}{6\Delta_+} +  \frac{\Omega^2_{-}}{6(\Delta_- - \omega_\text{HF})} \bigg) \cos^2(k_r x) + \frac{[g_0 \Xi(x,z)]^2}{\Delta_-} \cos^2(k_r y) \hat{a}^{\dagger} \hat{a} \Big] \hat{\psi}_{\downarrow} (\mbf{r}) \nonumber \\
H_{\mathrm{cavity}} &= -\Delta_c \hat{a}^{\dagger} \hat{a} \nonumber \\
H_{\mathrm{Raman}} &= \int d^3 \mbf{r} \Big[ \frac{\sqrt{3} g_0 \Xi(x,z) \Omega_+}{12 \Delta_+} \hat{\psi}^{\dagger}_{\uparrow} (\mbf{r}) \hat{\psi}_{\downarrow} (\mbf{r}) \hat{a}^{\dagger} \cos(k_r x) \cos(k_r y) + \frac{\sqrt{3} g_0 \Xi(x,z) \Omega_-}{12 \Delta_-} \hat{\psi}^{\dagger}_{\downarrow} (\mbf{r}) \hat{\psi}_{\uparrow} (\mbf{r}) \hat{a}^{\dagger} \cos(k_r x) \cos(k_r y) + \text{H.c.}\Big],
\end{align}
\end{widetext}
where we have separated out the longitudinal dependence of the cavity mode and $k_r = 2 \pi/\lambda$. In the regime of large cavity detuning $\Delta_c$, the dynamics of the cavity mode is  faster than other dynamics, and therefore we adiabatically eliminate the cavity mode to obtain an atom-only Hamiltonian. To do so, we define local spin operators
\begin{align}
\hat{\sigma}_{z} (\mbf{r}) &= \left[\hat{\psi}^{\dagger}_{\uparrow} (\mbf{r}) \hat{\psi}_{\uparrow} (\mbf{r}) - \hat{\psi}^{\dagger}_{\downarrow} (\mbf{r}) \hat{\psi}_{\downarrow}(\mbf{r}) \right]/2 \nonumber \\
\hat{\sigma}_x(\mbf{r}) &= \left[\hat{\psi}^\dagger_\uparrow(\mbf{r})\hat{\psi}_\downarrow(\mbf{r}) + \hat{\psi}^\dagger_\downarrow(\mbf{r})\hat{\psi}_\uparrow(\mbf{r})\right]/2.
\end{align}
The two cavity-assisted Raman couplings are set to have the same strength
\be\label{equalcoupling}
\frac{\sqrt{3} g_0 \Omega_-}{12 \Delta_-} = \frac{\sqrt{3} g_0 \Omega_+}{12 \Delta_+} \equiv \eta,
\ee
allowing the effective Hamiltonian to be written as 
\begin{align}
H_{\mrm{eff}} =& \int d^3 \mbf{r} d^3\mbf{r}' \frac{\eta^2}{\Delta_c} \Xi(x,z) \Xi(x',z') \times \nonumber \\ 
&\cos(k_r x) \cos(k_r x') \cos(k_r y) \cos(k_r y') \hat{\sigma}_x(\mbf{r}) \hat{\sigma}_x(\mbf{r}') \nonumber \\
&+ \int d^3 \mbf{r} (\hat{H}_{k} - \delta) \hat{\sigma}_z (\mbf{r}), 
\end{align}
where 
\begin{align}
\hat{H}_k =&-\frac{\hat{\mbf{p}}^2}{2 m} +  \bigg[ \frac{\Omega^2_{+}}{6(\Delta_+ + \omega_\text{HF})} +  \frac{\Omega^2_{-}}{6\Delta_-} \bigg] \cos^2(k_r x) \nonumber \\
& -\bigg[ \frac{\Omega^2_{+}}{6\Delta_+} +  \frac{\Omega^2_{-}}{6(\Delta_- - \omega_\text{HF})} \bigg] \cos^2(k_r x).
\end{align}
We have ignored the small Stark shift term proportional to $1/\Delta_{+,-}$ due to the  cavity field. Our system therefore realizes a transverse-field Ising model of the form 
\bea
H_\text{Ising} \propto && \sum J_{ij}\cos{k_rx_i}\cos{k_rx_j}\cos{k_ry_i}\cos{k_ry_j}\hat{\sigma}^i_x\hat{\sigma}^j_x  \nonumber\\ &&+ h\hat{\sigma}_z^i,
\eea
with direct spin-spin interaction mediated through the cavity mode.

\subsection{Mapping to the Dicke model}
To understand the threshold at which organization occurs, it is useful to map our system onto a Dicke model~\cite{Baumann2010,Nagy:2010dr}. The experiment begins with a  condensate in $\sd$ and  when the cavity-mediated Raman process  causes a spin flip, a momentum kick is also imparted onto the atoms.  Within the  single-recoil limit, the dynamics can be captured by two atomic modes
\begin{align}
\hat{\psi}_{\downarrow} &= \hat{c}_{\downarrow} \psi_0 \nonumber \\
\hat{\psi}_{\uparrow} &= \hat{c}_{\uparrow} \psi_1,
\end{align}
where, for simplicity,
\begin{align}
\psi_0 &= 1 \nonumber \\
\psi_1 &= 2 \cos(k_r x) \cos(k_r y).
\end{align}
Since the pump lattice potential is retained in the Hamiltonian, the differential Stark shift on $\psi_{\uparrow}$ and $\psi_{\downarrow}$ due to the lattice beams has been taken into account.
Taking advantage of the $\lambda = 2 \pi/k_r$ periodicity along both the pump and cavity direction, shifting the energy of the $c_\downarrow$ mode to zero, and  performing the integrals, the Hamiltonian is evaluated to be
\begin{widetext}
\begin{align}
H = &-\Delta_c \hat{a}^{\dagger} \hat{a} + \left\{ 2 \omega_r + \frac{3}{4}\left[\frac{\Omega^2_+}{6(\Delta_+ + \omega_\text{HF})}+ \frac{\Omega^2_-}{6 \Delta_-} \right] - \frac{1}{2} \left[ \frac{\Omega^2_+}{6 \Delta_+} + \frac{\Omega^2_-}{6(\Delta_- - \omega_\text{HF})} \right] -\delta \right\}\hat{c}^{\dagger}_{\uparrow} \hat{c}_{\uparrow} \nonumber \\
 &+\left[ \frac{\sqrt{3}}{24} \frac{g_0 \Omega_+}{\Delta_+} \hat{c}^{\dagger}_{\uparrow} \hat{c}_{\downarrow} + \frac{\sqrt{3}}{24} \frac{g_0 \Omega_-}{\Delta_-} \hat{c}^{\dagger}_{\downarrow} c_{\uparrow} \right] \Big(\hat{a}^{\dagger} + \hat{a} \Big) +\frac{g_0^2}{2 \Delta_-} \hat{c}^{\dagger}_{\downarrow} c_{\downarrow} \hat{a}^{\dagger} a + \frac{3 g_0^2}{4 \Delta_+} \hat{c}^{\dagger}_{\uparrow} c_{\uparrow} \hat{a}^{\dagger} a.
\end{align}
\end{widetext}
The bare energy of the $c_{\uparrow}$ mode is shifted due to the differential Stark shift
\be
\omega_{S} = \frac{3}{4}\bigg[\frac{\Omega^2_+}{6(\Delta_+ + \omega_\text{HF})}+ \frac{\Omega^2_-}{6 \Delta_-} \bigg] - \frac{1}{2} \bigg[ \frac{\Omega^2_+}{6 \Delta_+} + \frac{\Omega^2_-}{6(\Delta_- - \omega_\text{HF})} \bigg],
\ee
which is a dynamic quantity during the linear ramp  of the Raman-beams' power. The Raman detuning $\delta$ is therefore chosen such that the bare energy of the $\hat{c}_{\uparrow}$ mode is always positive during the experiment sequence.

As mentioned above in Eq.~\ref{equalcoupling},  the Raman couplings are chosen to be equal, and in anticipation of standard Dicke model notation~\cite{Kirton:2018vv}, we define this coupling as
\be
 \frac{\sqrt{3N}}{24} \frac{g_0 \Omega_+}{\Delta_+} = \frac{\sqrt{3N}}{24} \frac{g_0 \Omega_-}{\Delta_-}\equiv \eta_D,
\ee
where $N = \hat{c}^{\dagger}_{\uparrow} \hat{c}_{\uparrow} + \hat{c}^{\dagger}_{\downarrow} \hat{c}_{\downarrow}$ is the total number of  atoms.  We now define the collective pseudospin-1/2  operators
\begin{align}
\hat{J}_{z} &= \frac{1}{2} (\hat{c}^{\dagger}_{\uparrow} \hat{c}_{\uparrow} - \hat{c}^{\dagger}_{\downarrow} \hat{c}_{\downarrow}) \nonumber \\
\hat{J}_+ &= \hat{c}^{\dagger}_{\uparrow} \hat{c}_{\downarrow} \nonumber \\
\hat{J}_- &= \hat{c}^{\dagger}_{\downarrow} \hat{c}_{\uparrow},
\end{align}
where the $\hat{\mbf{J}}$ operate on the coupled pseudospin-1/2 spin-spatial degree of freedom.
The Hamiltonian can then be rewritten as
\begin{align}
H =&\bigg(-\Delta_c + \frac{N g_0^2}{2 \Delta_-} \bigg) \hat{a}^{\dagger} \hat{a} + (2 \omega_r + \omega_S - \delta) \hat{J}_{z} \nonumber \\
&+ \frac{\eta_D}{\sqrt{N}}(\hat{J}_+ + \hat{J}_-)(\hat{a}^{\dagger} + \hat{a}) \nonumber \\
 & +\frac{N (2 \omega_r + \omega_S - \delta)}{2} + \bigg(\frac{3 g_0^2}{4 \Delta_+} - \frac{g_0^2}{2 \Delta_-}\bigg) \hat{c}^{\dagger}_{\uparrow} c_{\uparrow} \hat{a}^{\dagger} \hat{a}.
\end{align}
The first term in the third line is simply a energy offset, while the second term can be ignored as long as the population of $\hat{c}^{\dagger}_{\uparrow}\hat{c}_{\uparrow} $ is small, which is consistent with the single-recoil limit. The Hamiltonian therefore realizes the Dicke model, and the usual threshold expression applies:
\be
\eta_c = \frac{1}{2} \sqrt{(-\Delta_c + N g_0^2/\Delta_{-})(2 \omega_r + \omega_S - \delta)}.
\ee

\subsection{Lattice calibration and Raman coupling balancing}
We calibrate the lattice depth of pump beams by performing Kapitza-Dirac diffraction of the BEC~\cite{Morsch2006} prepared in either $\su$ or $\sd$. This also allows us to characterize the differential Stark shift in the experiment. The retroreflection mirror shared by the pump beams is mounted on a translation stage. Measuring the lattice depth of the combined pump beams, we  adjust the translation stage to match the phases of the pump lattices at the position of the atoms. We note that the beat length of the two pump lattices (separated in optical frequency by 13.6~GHz) is $\sim$5~mm, much larger than the atomic cloud size; therefore, small mechanical fluctuations from the mirror mount will not cause the lattice to become out-of-phase at the atoms.

To match each beam's coupling strength, we linearly ramp up the $\Omega_+$ ($\Omega_-$) beam intensity for atoms prepared in $\sd$ ($\su$) while monitoring cavity emission. Above a certain pump strength, atoms are transferred to the other spin state with an accompanying brief cavity emission pulse. The critical Raman coupling strength at which this pulse occurs is given by: 
\be
\eta_{c,\mrm{single}} = \sqrt{\frac{\gamma \kappa}{2 N}\left[ 1 + \left( \frac{-\Delta_c + 2 \omega_r + \omega'_{S} - \delta}{\gamma + \kappa} \right)^2 \right]},
\ee
where $\kappa$ is the cavity decay rate and $\gamma$ is a phenomenological parameter describing the collective spin decay rate~\cite{Zhiqiang2018}. Since only one beam is involved, $\omega'_{S}$ denotes the differential Stark shift on $\su$ and $\sd$ due to a single pump beam. Matching the threshold for a single-beam spin-flip then balances the two Raman coupling processes. We perform the calibration with $\tilde\Delta_c = -1.4~\mrm{MHz}$. This is  significantly larger than  the two-beam Stark-shift contribution, $2 \omega_r + \omega'_{S} - \delta \approx 10~\mrm{kHz}$.  Therefore, the additional Stark shift from the simultaneous presence of both beams does not alter the matching condition considerably.

\end{document}